%% file: template_acm_ccs.tex
\documentclass[sigconf,nonacm]{acmart}

\usepackage{booktabs}
\usepackage{graphicx}
\usepackage{enumitem}

\makeatletter
\providecommand{\@LN@col}[1]{}
\providecommand{\@LN}[2]{}
\makeatother

\newcommand{\isacmccs}{}

\acmConference[]{}{}{}
\acmYear{}
\renewcommand\footnotetextcopyrightpermission[1]{}
\makeatletter\renewcommand{\@acmBooktitle}{}\makeatother
\let\oldendabstract\endabstract
\def\endabstract{\oldendabstract\maketitle\vspace{1.2em}}

\begin{document}

%
%

\title{Eroding Trust in Real Speech: A Large-Scale Study of Human Audio Deepfake Perception}

\author{Nicolas M. M\"uller}
\affiliation{%
  \institution{Fraunhofer AISEC \& Resemble AI}
  \country{Germany}
}
\email{nicolas.mueller@aisec.fraunhofer.de}
\author{Wei Herng Choong}
\affiliation{%
  \institution{Fraunhofer AISEC}
  \country{Germany}
}
\email{wei.herng.choong@aisec.fraunhofer.de}

\input{content}

\bibliographystyle{ACM-Reference-Format}
\bibliography{mybib}

\end{document}

%% file: content.tex
\begin{abstract}
    Audio deepfakes have improved rapidly recently, yet their effect on human trust in real speech remains unstudied.
    We present the largest listening study on audio deepfake perception to date, collecting \input{res/auto_gen/counts/new_rounds}\unskip judgments from \input{res/auto_gen/counts/new_users}\unskip participants across \input{res/auto_gen/counts/new_attacks}\unskip text-to-speech and voice conversion systems. Our central finding is a skepticism shift: compared to a 2021 baseline, human accuracy on fake samples barely changed (\input{res/auto_gen/counts/old_acc_fake}\unskip\% to \input{res/auto_gen/counts/new_acc_fake}\unskip\%), but accuracy on real samples dropped from \input{res/auto_gen/counts/old_acc_real}\unskip\% to \input{res/auto_gen/counts/new_acc_real}\unskip\%. Participants are not worse at detecting synthesis artifacts; rather, they increasingly distrust authentic speech. Samples generated by commercial and autoregressive language model systems proved hardest to detect (\input{res/auto_gen/counts/grp_commercial_hacc}\unskip-- \input{res/auto_gen/counts/grp_arlm_hacc}\unskip\%), while those from traditional seq2seq and flow-matching models remain easier to spot (\input{res/auto_gen/counts/grp_seq2seq_hacc}\unskip-- \input{res/auto_gen/counts/grp_flow_hacc}\unskip\%). An ML detector that served as a reference point maintained over \input{res/auto_gen/counts/new_acc_ml}\unskip\% accuracy across all conditions. 
    Our results suggest that the primary threat posed by modern deepfakes may not be mere deception, but the erosion of trust in genuine audio.
\end{abstract}

\section{Introduction}

The proliferation of text-to-speech (TTS) and voice conversion (VC) systems poses growing threats to trust in audio media. Deepfake audio has been used for CEO fraud~\cite{stupp2019fraudsters} and raises concerns for misinformation and evidence tampering~\cite{chesney2019deepfakes}. More recently, cloned voices have enabled cryptocurrency scams, romance fraud, and impersonation attacks at scale~\cite{fbi2026sf,deloitte2024genai}, including a \$25M deepfake video-call heist~\cite{gover2024arup}, AI-generated political robocalls~\cite{fcc2024robocall}, and widespread voice-cloning phone scams~\cite{mcafee2023voicescams}. While machine-learning (ML)-based detectors achieve high accuracy on known attacks~\cite{tak2021rawnet2}, they often fail to generalize to unseen TTS models~\cite{muller2024mlaad}. Human perception thus remains a critical, yet understudied, line of defense.

A 2021 study~\cite{muller2022human} evaluated human perception on \input{res/auto_gen/counts/old_attacks}\unskip attacks from ASVspoof 2019~\cite{wang2020asvspoof}, finding that \input{res/auto_gen/counts/old_users}\unskip participants achieved \input{res/auto_gen/counts/old_acc_all}\unskip\% accuracy. Since then, the TTS landscape has transformed: autoregressive language models like VALL-E~\cite{wang2023valle} and Bark~\cite{suno2023bark}, flow-matching systems like F5-TTS~\cite{chen2024f5tts}, as well as commercial APIs such as Resemble~AI~\cite{resembleai2024} produce speech that is increasingly difficult to distinguish from real recordings. No large-scale study has revisited how humans perceive this new generation of deepfakes.

\newpage
\textbf{Contribution.} We address this gap with the largest audio deepfake perception study to date: \input{res/auto_gen/counts/new_users}\unskip participants provided \input{res/auto_gen/counts/new_rounds}\unskip judgments across \input{res/auto_gen/counts/new_attacks}\unskip TTS and VC systems spanning \input{res/auto_gen/counts/new_groups}\unskip architecture families. We benchmark human performance against an ML detector, analyze the effect of demographics (age, IT skill, native language), test whether participants improve with practice, and compare directly to the 2021 baseline.

\noindent Our key findings are:
\begin{itemize}[nosep,leftmargin=*]
  \item Human accuracy on audio deepfakes remained nearly unchanged (\input{res/auto_gen/counts/old_acc_fake}\unskip\%$\to$\input{res/auto_gen/counts/new_acc_fake}\unskip\%), but accuracy on genuine audio dropped from \input{res/auto_gen/counts/old_acc_real}\unskip\% to \input{res/auto_gen/counts/new_acc_real}\unskip\%, suggesting growing skepticism towards recorded speech.
  \item Deepfakes generated by commercial and autoregressive language model (AR-LM) systems are hardest to detect (\input{res/auto_gen/counts/grp_commercial_hacc}\unskip--\input{res/auto_gen/counts/grp_arlm_hacc}\unskip\%), while those from traditional seq2seq and flow-matching models remain easiest (\input{res/auto_gen/counts/grp_seq2seq_hacc}\unskip--\input{res/auto_gen/counts/grp_flow_hacc}\unskip\%).
  \item Our baseline ML detector outperforms humans at \input{res/auto_gen/counts/new_acc_ml}\unskip\%, stable from the 2021 baseline of \input{res/auto_gen/counts/old_acc_ml}\unskip\%.
  \item Participants' ability to detect deepfakes improves with practice, as reflected in higher accuracy over successive rounds.
  \item Age and native language show little association with detection ability; self-rated experts (IT skill 5) score ${\sim}$4~pp higher than everyone else, a small but significant effect.
\end{itemize}
We release the anonymized dataset\footnote{\scriptsize\url{https://huggingface.co/datasets/mueller91/human-perception-audio-deepfake-2026}} and analysis code to support reproducibility.

\section{Related Work}

\textbf{Human deepfake perception.}
A meta-analysis of 56~studies (86,155~participants) found an overall detection accuracy across different data modality of only 55.5\%, with audio deepfakes detected best at 62\%~\cite{diel2024metaanalysis}. For video deepfakes, Groh et~al.~\cite{groh2022deepfake} showed that 15,016~participants performed comparably to ML models, while Cooke et~al.~\cite{cooke2025cointoss} found cross-modal detection near chance level.

\textbf{Audio-specific studies.}
M{\"u}ller et~al.~\cite{muller2022human} evaluated \input{res/auto_gen/counts/old_users}\unskip~participants on \input{res/auto_gen/counts/old_attacks}\unskip~attacks from ASVspoof~2019~LA~\cite{wang2020asvspoof} and reported \input{res/auto_gen/counts/old_acc_all}\unskip\% accuracy, with an ML detector reaching \input{res/auto_gen/counts/old_acc_ml}\unskip\%. Mai et~al.~\cite{mai2023warning} found similar rates (73\%) across English and Mandarin, with minimal benefit from training. Warren et~al.~\cite{warren2024better} demonstrated that humans and ML detectors make complementary errors and that preconceptions about accents and noise mislead human judges. San Segundo et~al.~\cite{sansegundo2025perception} showed that language and speaking style influence perception, with a general bias toward classifying samples as real. Critically, all existing audio studies evaluate at most \input{res/auto_gen/counts/old_attacks}\unskip~attacks from pre-2021 systems and do not cover the modern TTS architectures that are now widely deployed.
This study directly extends~\cite{muller2022human} by replicating the same game interface and active-learning sampling procedure, enabling a controlled comparison across a four-year gap in TTS development.

\textbf{TTS evolution.}
Seq2seq models such as Tacotron~2~\cite{shen2018tacotron2} gave way to end-to-end approaches (VITS~\cite{kim2021vits}), diffusion-based systems (Grad-TTS~\cite{popov2021gradtts}, StyleTTS~2~\cite{li2023styletts2}), autoregressive language models over codec tokens (VALL-E~\cite{wang2023valle}, Bark~\cite{suno2023bark}), and flow-matching architectures (F5-TTS~\cite{chen2024f5tts}, CosyVoice~\cite{du2024cosyvoice}). VALL-E~2~\cite{wang2024valle2} claims human parity in zero-shot TTS, and commercial APIs such as Resemble~AI~\cite{resembleai2024} make high-quality synthesis widely accessible. Human perception has not been evaluated against these systems at scale.

\textbf{ML-based detection.}
The ASVspoof challenge series~\cite{wang2020asvspoof,liu2023asvspoof2021,wang2025asvspoof5} has driven progress from end-to-end RawNet2~\cite{tak2021rawnet2} to pre-trained detectors built on Wav2Vec~2.0~\cite{tak2022wav2vec} and AASIST~\cite{jung2022aasist}. However, ML detectors generalize poorly to unseen attacks~\cite{muller2022generalize}, making human perception a complementary and still poorly understood line of defense.
\begin{table}[t]
  \centering
  \caption{Overview of the 2021 and 2026 listening studies. Accuracy is reported with 95\% bootstrap CI.}
  \label{tab:overall}
  \input{tab/auto_gen/overall}
\end{table}

\section{Study Design}
In this section, we describe our study setup, including the web platform and game procedure, the composition of the audio corpus, the ML deepfake detector that serves as a reference for human performance, and the filtering criteria for constructing the final analysis set. 

\subsection{Web platform and procedure}

We host a publicly accessible listening game online\footnote{\scriptsize\url{https://deepfake-total.com/spot_the_audio_deepfake}}.
Before playing, each participant answered three demographic questions: age, self-rated IT skill (Likert scale, 1=novice to 5=expert), and whether English is their native language.
No registration or account is required; we collect no data beyond what users provide voluntarily.

The listening game consists of individual rounds, of which the user can play as many as they please. In each round, the system presents a single audio clip.
With probability $p{=}0.5$ the clip is bona fide; otherwise a fake attack is selected using a weighted active-learning scheme~\cite{muller2022human}: the sampling weight for attack~$i$ is $w_i = 1 - \mathrm{acc}_i / (1 + \varepsilon)$, where $\mathrm{acc}_i$ is the running human accuracy on that attack, so that trivially detectable systems are presented less often. $\varepsilon$ is a small constant ensuring a non-zero sampling probability.
Participants may replay the clip as often as they wish before classifying it as \emph{real} or \emph{fake}.

After the decision is made, the interface reveals both the ground truth and the ML detector's prediction, providing participants with immediate feedback.
The platform is mobile-responsive and all audio is in English.
The interface and procedure follow~\cite{muller2022human}.

\subsection{Audio data}

Bona fide samples are drawn from three sources: LJSpeech~\cite{ljspeech17}, the In-The-Wild corpus~\cite{muller2022generalize}, and ASVspoof 5~\cite{wang2025asvspoof5}.
Fake samples are largely drawn from ASVspoof 5 and the English subset of MLAAD~\cite{muller2024mlaad}. They cover \input{res/auto_gen/counts/new_attacks}\unskip TTS and VC systems across \input{res/auto_gen/counts/new_groups}\unskip architecture families (c.f.\ Figure~\ref{fig:per_model}):
\begin{itemize}[nosep,leftmargin=*]
  \item \textbf{\emph{Seq2Seq}} -- encoder-decoder, e.g.\ Tacotron~2~\cite{shen2018tacotron2}
  \item \textbf{\emph{VITS}} -- VAE + flow + GAN~\cite{kim2021vits}
  \item \textbf{\emph{XTTS}} -- GPT-based multi-speaker TTS with VITS decoder
  \item \textbf{\emph{Flow}} -- flow-matching, e.g.\ F5-TTS~\cite{chen2024f5tts}, CosyVoice~\cite{du2024cosyvoice}
  \item \textbf{\emph{Diffusion}} -- e.g.\ Grad-TTS~\cite{popov2021gradtts}, StyleTTS~2~\cite{li2023styletts2}
  \item \textbf{\emph{AR-LM}} -- autoregressive language models over codec tokens, e.g.\ VALL-E~\cite{wang2023valle}, Bark~\cite{suno2023bark}, Chatterbox~\cite{chatterbox2025}
  \item \textbf{\emph{VC}} -- voice conversion, e.g.\ RVC, OpenVoice~V2~\cite{qin2024openvoice}
  \item \textbf{\emph{Commercial}} -- proprietary APIs such as ElevenLabs~\cite{elevenlabs2024} and Resemble~AI~\cite{resembleai2024}
  \item \textbf{\emph{ASVspoof~5}} -- attacks from~\cite{wang2025asvspoof5}
\end{itemize}
This represents a substantial expansion over the 2021 study~\cite{muller2022human}, which used only \input{res/auto_gen/counts/old_attacks}\unskip attacks from ASVspoof 2019 LA.

\begin{figure}[t]
  \centering
  \includegraphics[width=0.9\columnwidth]{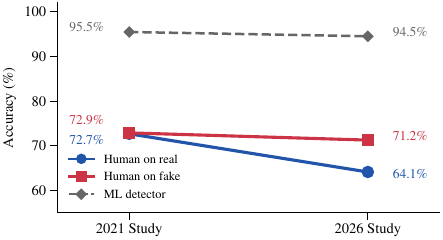}
  \caption{Human accuracy on real vs.\ fake samples in 2021 and 2026. Accuracy on fakes is stable (\protect\input{res/auto_gen/counts/old_acc_fake}\unskip\%$\to$\protect\input{res/auto_gen/counts/new_acc_fake}\unskip\%), but accuracy on real samples dropped sharply (\protect\input{res/auto_gen/counts/old_acc_real}\unskip\%$\to$\protect\input{res/auto_gen/counts/new_acc_real}\unskip\%), indicating growing skepticism. ML detector average accuracy (dashed) remains above 93\% in both studies.}
  \label{fig:overall_accuracy}
\end{figure}

\subsection{ML classifier}

An in-house ML deepfake detector runs alongside the human game as a reference point.
The model combines Wav2Vec~2.0 features~\cite{baevski2020wav2vec} with an AASIST~\cite{jung2022aasist} back-end, trained on a mix of public and internal data.
Its predictions are logged for every round, enabling direct human--machine comparison.

\subsection{Filtering}

We exclude participants who completed fewer than 5~rounds and attacks with fewer than 10~judgments, retaining \input{res/auto_gen/counts/new_users}\unskip~participants and \input{res/auto_gen/counts/new_attacks}\unskip~attacks for the 2026 study (Table~\ref{tab:overall}).

\begin{figure*}[!ht]
  \centering
  \includegraphics[width=\textwidth]{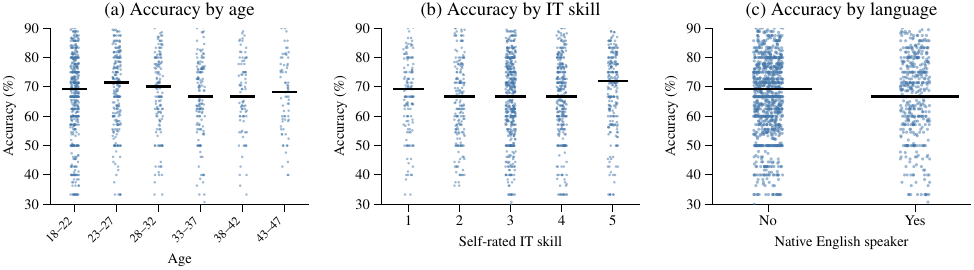}
  \caption{Demographic analysis of the 2026 study. (a)~Accuracy by age bracket (18--49); each dot is one participant, horizontal lines show medians. (b)~Accuracy by self-rated IT skill (1=novice, 5=expert): levels 1--4 are indistinguishable (median ${\sim}$67\%), but self-rated experts (skill=5) score ${\sim}$4~pp higher (median 72\%), a small but statistically significant effect. (c)~Accuracy by native English speaker status.}
  \label{fig:demographics}
\end{figure*}

\begin{figure}[!ht]
  \centering
  \includegraphics[width=\columnwidth]{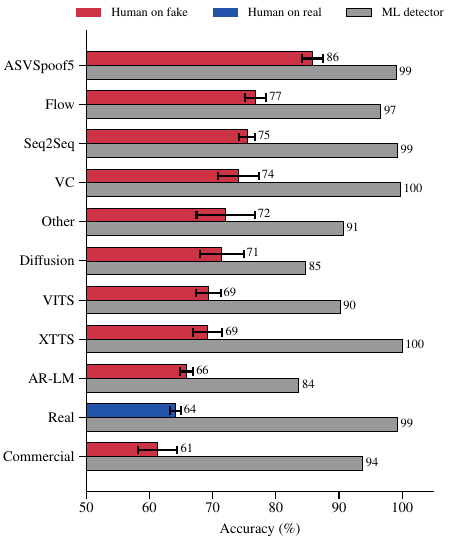}
  \caption{Human (colored) and ML detector (grey) accuracy per TTS architecture group (red) and on real samples (blue), sorted by human accuracy. Commercial APIs and AR-LM systems generate samples that are hardest for humans to detect. Notably, human accuracy on real audio ranks below most fake categories, visually illustrating the skepticism shift. Error bars show 95\% confidence interval (CI).}
  \label{fig:accuracy_by_group}
\end{figure}

\section{Results}

Table~\ref{tab:overall} summarizes both studies. The 2026 study is substantially larger than the 2021 study: \input{res/auto_gen/counts/new_users}\unskip~participants provided \input{res/auto_gen/counts/new_rounds}\unskip~judgments across \input{res/auto_gen/counts/new_attacks}\unskip~attacks from \input{res/auto_gen/counts/new_groups}\unskip~architecture families, compared to \input{res/auto_gen/counts/old_users}\unskip~participants and \input{res/auto_gen/counts/old_attacks}\unskip~attacks in 2021.

The central finding is an asymmetric shift in human accuracy (Figure~\ref{fig:overall_accuracy}). On fake samples, accuracy barely changed: \input{res/auto_gen/counts/old_acc_fake}\unskip\%$\to$\input{res/auto_gen/counts/new_acc_fake}\unskip\%. On real samples, however, accuracy dropped from \input{res/auto_gen/counts/old_acc_real}\unskip\% to \input{res/auto_gen/counts/new_acc_real}\unskip\%, an \input{res/auto_gen/counts/delta_acc_real}\unskip percentage point decline. These results suggest that participants are not worse at hearing synthesis artifacts; rather, they have become more skeptical of all audio, increasingly classifying genuine recordings as fake. The ML detector remained stable at \input{res/auto_gen/counts/old_acc_ml}\unskip\% (2021) and \input{res/auto_gen/counts/new_acc_ml}\unskip\% (2026), outperforming humans by more than 20 percentage points in both periods.


\subsection{Which architectures fool humans?}

\begin{figure*}[p]
  \centering
  \resizebox{!}{0.92\textheight}{\includegraphics{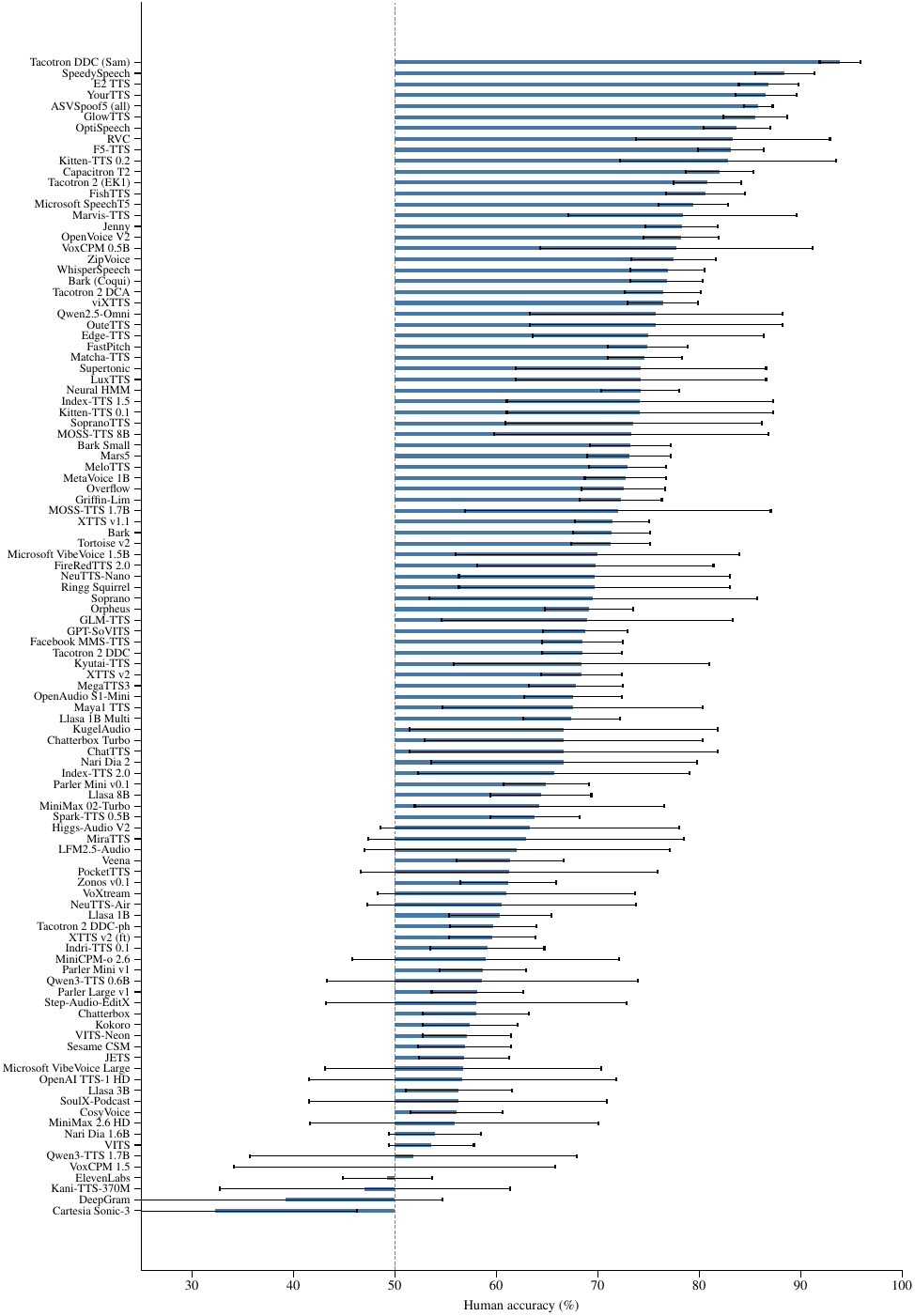}}
  \caption{Human accuracy for each individual TTS/VC system (fake samples only, minimum 10 judgments), sorted by descending accuracy. Error bars show 90\% CI.}
  \label{fig:per_model}
\end{figure*}

Figure~\ref{fig:accuracy_by_group} 
breaks down accuracy by TTS architecture family. Fake samples generated by commercial APIs are the hardest for humans to detect (\input{res/auto_gen/counts/grp_commercial_hacc}\unskip\%), followed by samples from AR-LM systems (\input{res/auto_gen/counts/grp_arlm_hacc}\unskip\%). These two families are also the most practically relevant: commercial APIs are widely accessible, and AR-LM models dominate the open-source landscape. Together, they account for \input{res/auto_gen/counts/grp_modern_nmodels}\unskip~models and \input{res/auto_gen/counts/grp_modern_nsamples}\unskip~samples, making them the largest group in our study. On the other hand, the samples generated by traditional seq2seq systems (\input{res/auto_gen/counts/grp_seq2seq_hacc}\unskip\%), flow-matching models (\input{res/auto_gen/counts/grp_flow_hacc}\unskip\%), and ASVspoof~5 (\input{res/auto_gen/counts/grp_asvspoof5_hacc}\unskip\%) remain substantially easier to detect. The ML detector maintains above \input{res/auto_gen/counts/grp_ml_min}\unskip\% accuracy across all architecture families and exceeds 99\% on 4 of the 10 families. AR-LM and diffusion are the only families in which ML accuracy drops below 90\% (\input{res/auto_gen/counts/grp_arlm_mlacc}\unskip\% and \input{res/auto_gen/counts/grp_diffusion_mlacc}\unskip\%, respectively), making them the most challenging for automated detection. See Figure~\ref{fig:per_model} for a detailed, per-model analysis.

\subsection{Demographics and learning}

\begin{figure}[t]
  \centering
  \includegraphics[width=\columnwidth]{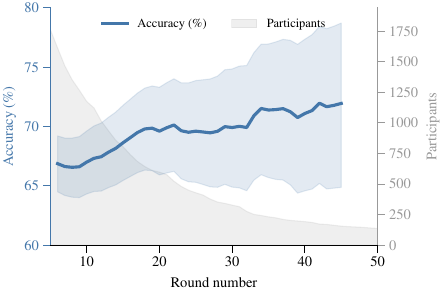}
  \caption{Learning effect: accuracy by round number (blue, 10-round moving average with 95\% CI, left axis) and number of active participants (grey, right axis). Accuracy plateaus after ${\sim}$20 rounds; participant count drops steadily as users stop playing.}
  \label{fig:learning}
\end{figure}

Self-reported demographics had limited effect on detection ability (Figure~\ref{fig:demographics}). Accuracy showed no meaningful trend with age, and native English speakers achieved no significant advantage over non-native speakers. Self-rated IT skill is mostly flat: levels 1--4 are indistinguishable (median ${\sim}$67\%), but self-rated experts (skill=5) score ${\sim}$4~pp higher (median 72\%), a small but statistically significant effect (Mann-Whitney $p{<}0.001$). These findings contrast with the 2021 study, where native speakers showed a clearer advantage and age had a negative correlation. Regarding learning effects (Figure~\ref{fig:learning}), participants in their first five rounds achieved only \input{res/auto_gen/counts/new_acc_first5}\unskip\% accuracy, compared to \input{res/auto_gen/counts/new_acc_after15}\unskip\% for those who had played more than 15~rounds. Accuracy improved steadily over the first ${\sim}$20~rounds, after which it plateaued regardless of further exposure, even as participant count dropped from \input{res/auto_gen/counts/new_users}\unskip to \input{res/auto_gen/counts/new_users_round50}\unskip by round~50. This mirrors the 2021 study, where immediate feedback likewise helped participants calibrate quickly but offered diminishing returns beyond the initial learning phase.

\section{Discussion}

\textbf{Skepticism as a societal risk.}
The key result is not that humans fail to detect fakes, as they perform comparably to 2021 on fake samples, but that they increasingly distrust real audio. This ``skepticism shift'' has direct implications beyond deepfake detection: if listeners routinely doubt authentic recordings, the evidentiary value of audio in journalism, legal proceedings, and personal communication erodes regardless of whether actual deepfakes are involved. Therefore, awareness and media-literacy campaigns should address over-skepticism alongside deepfake detection training.

\textbf{Architecture as threat model.}
Commercial APIs and AR-LM systems, which are among the most accessible synthesis technologies, produce fake samples that are also the hardest for humans to detect. The overall stability of human detection accuracy on fake samples (\input{res/auto_gen/counts/old_acc_fake}\unskip\%$\to$\input{res/auto_gen/counts/new_acc_fake}\unskip\%) masks a composition effect: for commercial and AR-LM systems, human accuracy drops to \input{res/auto_gen/counts/grp_modern_hacc}\unskip\%, while fake samples from legacy architectures are detected at \input{res/auto_gen/counts/grp_legacy_hacc}\unskip\%. This creates a concerning alignment: the systems most likely to be misused are precisely those that best evade human judgment. Traditional seq2seq and flow-matching systems, though still actively developed, produce speech that humans detect more reliably, suggesting that their acoustic characteristics differ in perceptually salient ways.

\textbf{Human--ML complementarity.}
Our ML detector maintained above \input{res/auto_gen/counts/grp_ml_min}\unskip\% accuracy even on categories where humans dropped to \input{res/auto_gen/counts/grp_commercial_hacc}\unskip\%, but ML detectors are known to generalize poorly to unseen attacks~\cite{muller2022generalize}. Meanwhile, AR-LM and diffusion systems proved challenging for the ML detector (\input{res/auto_gen/counts/grp_arlm_mlacc}\unskip\% and \input{res/auto_gen/counts/grp_diffusion_mlacc}\unskip\%, respectively), indicating that no single defense is sufficient. Combining human and machine judgment, for instance, by flagging ML-uncertain samples for human review, could exploit their complementary error patterns.

\textbf{Limitations.}
Participants were self-selected through a web-based game, skewing toward younger users (cf. Figure~\ref{fig:demographics}). Audio quality varied with participants' playback equipment and browser compression. The study is in English-only, and participation is open and anonymous, so we cannot control for users who participated in both 2021 and 2026 studies. The active-learning sampling scheme, while improving coverage of difficult attacks, produces uneven per-attack sample counts.

\section{Conclusion}

We presented the largest listening study on audio deepfake perception to date. Our central finding is a skepticism shift: human accuracy on fake samples has remained stable since 2021, but accuracy on real samples dropped by \input{res/auto_gen/counts/delta_acc_real}\unskip percentage points, suggesting that the primary effect of improved deepfakes is erosion of trust in authentic speech rather than inability to detect synthesis. Commercial and AR-LM systems, the most widely deployed architectures, are hardest for humans to detect, while ML detectors remain effective across all conditions. Future work should extend this paradigm to multilingual settings, track longitudinal changes in listener behavior, and explore human--ML collaborative detection.

\ifdefined\isacmccs
\section*{Ethical Considerations}
This study was conducted via a publicly accessible web-based listening game. Participation is voluntary and requires no registration or account. We collect three coarse self-reported attributes (age bracket, IT skill level on a 1--5 scale, and whether English is the participant's native language). No directly identifiable information such as names, email addresses, or IP addresses is recorded. For analysis, all attributes are aggregated into group-level statistics. Participants receive immediate feedback after each round. All audio samples are drawn from published corpora or publicly available TTS systems. This study involves only anonymous, voluntary interaction with publicly available audio stimuli, placing it below the threshold for formal ethical review under standard institutional guidelines for minimal-risk research.

\section*{Open Science}
The listening platform is publicly accessible\footnote{\scriptsize\url{https://deepfake-total.com/spot_the_audio_deepfake}}.
All audio samples are drawn from publicly available corpora: LJSpeech~\cite{ljspeech17}, In-The-Wild~\cite{muller2022generalize}, ASVspoof~5~\cite{wang2025asvspoof5}, and MLAAD~\cite{muller2024mlaad}.
The anonymized per-round dataset is released on Hugging Face\footnote{\scriptsize\url{https://huggingface.co/datasets/mueller91/human-perception-audio-deepfake-2026}}. The dataset contains \input{res/auto_gen/counts/new_rounds}\unskip~rows from \input{res/auto_gen/counts/new_users}\unskip~participants across \input{res/auto_gen/counts/new_attacks}\unskip~attacks. Each row records the audio filename, attack ID, ground truth, as well as human and ML detector decisions. Self-reported demographic attributes (age, IT skill, native language) are excluded to prevent potential re-identification from response patterns. Table~\ref{tab:dataset} describes the released fields.

\begin{table}[h]
  \centering
  \caption{Fields in the released dataset.}
  \label{tab:dataset}
  \small
  \begin{tabular}{@{}ll@{}}
    \toprule
    Field & Description \\
    \midrule
    \texttt{uid} & Anonymized participant identifier \\
    \texttt{rounds\_played} & Round number for this participant \\
    \texttt{filename} & Audio filename (from public corpora) \\
    \texttt{attack\_id} & TTS/VC system or real-audio source \\
    \texttt{true\_label} & Ground truth (\texttt{real} / \texttt{fake}) \\
    \texttt{user\_decision} & Participant's response (\texttt{real} / \texttt{fake}) \\
    \texttt{ml\_decision} & ML detector's prediction (\texttt{real} / \texttt{fake}) \\
    \bottomrule
  \end{tabular}
\end{table}
\fi

%% file: res/auto_gen/counts/new_rounds.tex
35,532

%% file: res/auto_gen/counts/new_users.tex
1,768

%% file: res/auto_gen/counts/new_attacks.tex
138

%% file: res/auto_gen/counts/old_acc_fake.tex
72.9

%% file: res/auto_gen/counts/new_acc_fake.tex
71.2

%% file: res/auto_gen/counts/old_acc_real.tex
72.7

%% file: res/auto_gen/counts/new_acc_real.tex
64.1

%% file: res/auto_gen/counts/grp_commercial_hacc.tex
61.3

%% file: res/auto_gen/counts/grp_arlm_hacc.tex
65.9

%% file: res/auto_gen/counts/grp_seq2seq_hacc.tex
75.4

%% file: res/auto_gen/counts/grp_flow_hacc.tex
76.8

%% file: res/auto_gen/counts/new_acc_ml.tex
94.5

%% file: res/auto_gen/counts/old_attacks.tex
13

%% file: res/auto_gen/counts/old_users.tex
472

%% file: res/auto_gen/counts/old_acc_all.tex
72.8

%% file: res/auto_gen/counts/new_groups.tex
10

%% file: res/auto_gen/counts/old_acc_ml.tex
95.5

%% file: tab/auto_gen/overall.tex
\small
\begin{tabular}{@{}lrr@{}}
\toprule
& 2021 & 2026 \\
\midrule
  Users & 472 & 1,768 \\
  Rounds & 14,912 & 35,532 \\
  Attacks & 13 & 138 \\
  Human accuracy & 72.8\% $\pm$ 0.7 & 68.7\% $\pm$ 0.5 \\
  ~~on real samples & 72.7\% & 64.1\% \\
  ~~on fake samples & 72.9\% & 71.2\% \\
  ML accuracy & 95.5\% & 94.5\% \\
\bottomrule
\end{tabular}

%% file: res/auto_gen/counts/delta_acc_real.tex
8.6

%% file: res/auto_gen/counts/grp_modern_nmodels.tex
65

%% file: res/auto_gen/counts/grp_modern_nsamples.tex
8,929

%% file: res/auto_gen/counts/grp_asvspoof5_hacc.tex
85.9

%% file: res/auto_gen/counts/grp_ml_min.tex
84

%% file: res/auto_gen/counts/grp_arlm_mlacc.tex
83.7

%% file: res/auto_gen/counts/grp_diffusion_mlacc.tex
84.7

%% file: res/auto_gen/counts/new_acc_first5.tex
67.0

%% file: res/auto_gen/counts/new_acc_after15.tex
71.0

%% file: res/auto_gen/counts/new_users_round50.tex
136

%% file: res/auto_gen/counts/grp_modern_hacc.tex
65.4

%% file: res/auto_gen/counts/grp_legacy_hacc.tex
74.9

%% file: mybib.bib
@inproceedings{muller2022human,
  author={Nicolas M. M{\"u}ller and Karla Pizzi and Jennifer Williams},
  title={Human Perception of Audio Deepfakes},
  booktitle={Proc. 1st International Workshop on Deepfake Detection for Audio Multimedia (DDAM)},
  year={2022},
  pages={85--91},
  doi={10.1145/3552466.3556531}
}

@inproceedings{muller2022generalize,
  author={Nicolas M. M{\"u}ller and Pavel Czempin and Franziska Diekmann and Adam Froghyar and Konstantin B{\"o}ttinger},
  title={Does Audio Deepfake Detection Generalize?},
  booktitle={Proc. Interspeech},
  year={2022},
  pages={2783--2787}
}

@article{wang2020asvspoof,
  author={Xin Wang and Junichi Yamagishi and Massimiliano Todisco and H{\'e}ctor Delgado and Andreas Nautsch and Nicholas Evans and others},
  title={{ASV}spoof 2019: A Large-Scale Public Database of Synthesized, Converted and Replayed Speech},
  journal={Computer Speech \& Language},
  volume={64},
  year={2020},
  pages={101114}
}

@article{liu2023asvspoof2021,
  author={Xuechen Liu and Xin Wang and Md Sahidullah and Jose Patino and H{\'e}ctor Delgado and Tomi Kinnunen and Massimiliano Todisco and Junichi Yamagishi and Nicholas Evans and Andreas Nautsch and Kong Aik Lee},
  title={{ASV}spoof 2021: Towards Spoofed and Deepfake Speech Detection in the Wild},
  journal={IEEE/ACM Transactions on Audio, Speech and Language Processing},
  year={2023}
}

@article{wang2025asvspoof5,
  author={Xin Wang and H{\'e}ctor Delgado and Hemlata Tak and Jee-weon Jung and others},
  title={{ASV}spoof 5: Design, Collection and Validation of Resources for Spoofing, Deepfake, and Adversarial Attack Detection Using Crowdsourced Speech},
  journal={Computer Speech \& Language},
  year={2025}
}

@misc{ljspeech17,
  author={Keith Ito and Linda Johnson},
  title={The {LJ} Speech Dataset},
  year={2017},
  howpublished={\url{https://keithito.com/LJ-Speech-Dataset/}}
}

@inproceedings{shen2018tacotron2,
  author={Jonathan Shen and Ruoming Pang and Ron J. Weiss and Mike Schuster and Navdeep Jaitly and Zongheng Yang and others},
  title={Natural {TTS} Synthesis by Conditioning {W}ave{N}et on Mel Spectrogram Predictions},
  booktitle={Proc. ICASSP},
  year={2018},
  pages={4779--4783}
}

@inproceedings{kim2021vits,
  author={Jaehyeon Kim and Jungil Kong and Juhee Son},
  title={Conditional Variational Autoencoder with Adversarial Learning for End-to-End Text-to-Speech},
  booktitle={Proc. ICML},
  year={2021},
  pages={5530--5540}
}

@article{wang2023valle,
  author={Chengyi Wang and Sanyuan Chen and Yu Wu and Ziqiang Zhang and Long Zhou and Shujie Liu and others},
  title={{VALL-E}: Neural Codec Language Models are Zero-Shot Text to Speech Synthesizers},
  journal={arXiv preprint arXiv:2301.02111},
  year={2023}
}

@article{wang2024valle2,
  author={Sanyuan Chen and Shujie Liu and Long Zhou and Yanqing Liu and Xu Tan and Jinyu Li and Sheng Zhao and Yao Qian and Furu Wei},
  title={{VALL-E} 2: Neural Codec Language Models are Human Parity Zero-Shot Text to Speech Synthesizers},
  journal={arXiv preprint arXiv:2406.05370},
  year={2024}
}

@misc{suno2023bark,
  author={{Suno AI}},
  title={Bark: Text-to-Audio Model},
  year={2023},
  note={\url{https://github.com/suno-ai/bark}}
}

@inproceedings{chen2024f5tts,
  author={Yushen Chen and Zhikang Niu and Ziyang Ma and Keqi Deng and Chunhui Wang and Jian Zhao and others},
  title={{F5-TTS}: A Fairytaler that Fakes Fluent and Faithful Speech with Flow Matching},
  booktitle={Proc. ACL},
  year={2025},
  pages={6255--6271}
}

@inproceedings{popov2021gradtts,
  author={Vadim Popov and Ivan Vovk and Vladimir Gogoryan and Tasnima Sadekova and Mikhail Kudinov},
  title={{Grad-TTS}: A Diffusion Probabilistic Model for Text-to-Speech},
  booktitle={Proc. ICML},
  year={2021}
}

@inproceedings{li2023styletts2,
  author={Yinghao Aaron Li and Cong Han and Vinay S. Raghavan and Gavin Mischler and Nima Mesgarani},
  title={{StyleTTS} 2: Towards Human-Level Text-to-Speech through Style Diffusion and Adversarial Training with Large Speech Language Models},
  booktitle={Proc. NeurIPS},
  year={2023}
}

@article{du2024cosyvoice,
  author={Zhihao Du and Qian Chen and others},
  title={{CosyVoice}: A Scalable Multilingual Zero-shot Text-to-speech Synthesizer Based on Supervised Semantic Tokens},
  journal={arXiv preprint arXiv:2407.05407},
  year={2024}
}

@inproceedings{tak2021rawnet2,
  author={Hemlata Tak and Jose Patino and Massimiliano Todisco and Andreas Nautsch and Nicholas Evans and Anthony Larcher},
  title={End-to-End Anti-Spoofing with {R}aw{N}et2},
  booktitle={Proc. ICASSP},
  year={2021},
  pages={6369--6373}
}

@inproceedings{jung2022aasist,
  author={Jee-weon Jung and Hee-Soo Heo and Hemlata Tak and Hye-jin Shim and Joon Son Chung and Bong-Jin Lee and Ha-Jin Yu and Nicholas Evans},
  title={{AASIST}: Audio Anti-Spoofing Using Integrated Spectro-Temporal Graph Attention Networks},
  booktitle={Proc. ICASSP},
  year={2022},
  pages={6367--6371}
}

@inproceedings{tak2022wav2vec,
  author={Hemlata Tak and Massimiliano Todisco and Xin Wang and Jee-weon Jung and Junichi Yamagishi and Nicholas Evans},
  title={Automatic Speaker Verification Spoofing and Deepfake Detection Using {Wav2Vec} 2.0 and Data Augmentation},
  booktitle={Proc. Speaker Odyssey},
  year={2022}
}

@inproceedings{muller2024mlaad,
  author={Nicolas M. M{\"u}ller and Piotr Kawa and Wei Herng Choong and others},
  title={{MLAAD}: The Multi-Language Audio Anti-Spoofing Dataset},
  booktitle={Proc. IJCNN},
  year={2024},
  doi={10.1109/IJCNN60899.2024.10650962}
}

@article{baevski2020wav2vec,
  title={wav2vec 2.0: A framework for self-supervised learning of speech representations},
  author={Baevski, Alexei and Zhou, Yuhao and Mohamed, Abdelrahman and Auli, Michael},
  journal={Advances in neural information processing systems},
  volume={33},
  pages={12449--12460},
  year={2020}
}

@article{mai2023warning,
  author={Khai Tinh Mai and Sergi D. Bray and Toby O. Davies and Lewis D. Griffin},
  title={Warning: Humans Cannot Reliably Detect Speech Deepfakes},
  journal={PLOS ONE},
  year={2023}
}

@article{groh2022deepfake,
  author={Matthew Groh and Ziv Epstein and Chaz Firestone and Rosalind Picard},
  title={Deepfake Detection by Human Crowds, Machines, and Machine-Informed Crowds},
  journal={Proceedings of the National Academy of Sciences},
  volume={119},
  number={1},
  year={2022}
}

@inproceedings{warren2024better,
  author={Kevin Warren and Tyler Tucker and Anna Crowder and Daniel Olszewski and Allison Lu and Caroline Fedele and Magdalena Pasternak and Seth Layton and Kevin Butler and Carrie Gates and Patrick Traynor},
  title={Better Be Computer or {I'm} Dumb: A Large-Scale Evaluation of Humans as Audio Deepfake Detectors},
  booktitle={Proc. ACM CCS},
  year={2024},
  doi={10.1145/3658644.3670325}
}

@article{cooke2025cointoss,
  author={Di Cooke and Abigail Edwards and Sophia Barkoff and Kathryn Kelly},
  title={As Good as a Coin Toss: Human Detection of {AI}-Generated Content},
  journal={Communications of the ACM},
  volume={68},
  number={10},
  year={2025}
}

@article{sansegundo2025perception,
  author={Eugenia {San Segundo} and Aurora L{\'o}pez-Jare{\~n}o and Xin Wang and Junichi Yamagishi},
  title={Human Perception of Audio Deepfakes: The Role of Language and Speaking Style},
  journal={arXiv preprint arXiv:2512.09221},
  year={2025}
}

@article{diel2024metaanalysis,
  author={Alexander Diel and Tania Lalgi and Isabel C. Schr{\"o}ter and Karl F. MacDorman and Martin Teufel and Alexander B{\"a}uerle},
  title={Human Performance in Detecting Deepfakes: A Systematic Review and Meta-Analysis of 56 Papers},
  journal={Computers in Human Behavior Reports},
  volume={16},
  year={2024},
  doi={10.1016/j.chbr.2024.100499}
}

@misc{fbi2026sf,
  author={{FBI San Francisco}},
  title={FBI San Francisco Warns Romance Scams Increasing Across the Bay Area This Valentine’s Day},
  year={2026},
  howpublished={https://www.fbi.gov/contact-us/field-offices/sanfrancisco/fbi-san-francisco-warns-romance-scams-increasing-across-the-bay-area-this-valentines-day}
}

@misc{deloitte2024genai,
  author={{Deloitte Center for Financial Services}},
  title={Generative {AI} Is Expected to Magnify the Risk of Deepfakes and Other Fraud in Banking},
  year={2024},
  howpublished={\url{https://www2.deloitte.com/us/en/insights/industry/financial-services/financial-services-industry-predictions/2024/deepfake-banking-fraud-risk-on-the-rise.html}}
}

@article{stupp2019fraudsters,
  author={Catherine Stupp},
  title={Fraudsters Used {AI} to Mimic {CEO}'s Voice in Unusual Cybercrime Case},
  journal={The Wall Street Journal},
  year={2019},
  month={Aug}
}

@article{chesney2019deepfakes,
  author={Robert Chesney and Danielle Citron},
  title={Deep Fakes: A Looming Challenge for Privacy, Democracy, and National Security},
  journal={California Law Review},
  volume={107},
  pages={1753--1820},
  year={2019}
}

@misc{resembleai2024,
  author={{Resemble AI}},
  title={{Resemble AI} Speech Synthesis {API}},
  year={2024},
  note={\url{https://www.resemble.ai}}
}

@misc{elevenlabs2024,
  author={{ElevenLabs}},
  title={{ElevenLabs Text to Speech API}},
  year={2024},
  howpublished={\url{https://elevenlabs.io}}
}

@misc{chatterbox2025,
  author={{Resemble AI}},
  title={{Chatterbox TTS}},
  year={2025},
  howpublished={\url{https://github.com/resemble-ai/chatterbox}}
}

@article{gover2024arup,
  author={Daniel Gover},
  title={Finance worker pays out \$25 million after video call with deepfake `chief financial officer'},
  journal={CNN},
  year={2024},
  month={Feb}
}

@misc{fcc2024robocall,
  author={{Federal Communications Commission}},
  title={Proposed \$6 Million Fine Against Political Consultant Who Used {AI}-Generated Deepfake Robocalls},
  year={2024},
  howpublished={\url{https://docs.fcc.gov/public/attachments/DOC-402762A1.pdf}}
}

@misc{mcafee2023voicescams,
  author={{McAfee}},
  title={Beware the Artificial Impostor: {A McAfee} Study on the Rise of {AI} Scams},
  year={2023},
  howpublished={\url{https://www.mcafee.com/learn/a-guide-to-deepfake-scams-and-ai-voice-spoofing/}}
}

@article{qin2024openvoice,
  author={Qin, Zengyi and Zhao, Wenliang and Yu, Xumin and Sun, Xin},
  title={{OpenVoice}: Versatile Instant Voice Cloning},
  journal={arXiv preprint arXiv:2312.01479},
  year={2024}
}
